\documentstyle[epsf]{aipproc}
\begin{document}

\title{Hot Evolved Stars in the Centers of M~31 and M~32}

\author{Thomas M. Brown $^{ \dagger}$, Henry C. Ferguson$^{ \dagger\dagger}$, S. Adam Stanford$^{\dagger\dagger\dagger}$, Jean-Michel Deharveng$^*$, and Arthur F. Davidsen$^{**}$}

\address{
$^{ \dagger}$NASA/GSFC, Code 681, Greenbelt, MD 20771 \\ 
$^{ \dagger\dagger}$STScI, 3700 San Martin Drive, Baltimore, MD 21218 \\
$^{\dagger\dagger\dagger}$IGPP, LLNL, 7000 East Avenue, Livermore, CA  94550\\ 
$^*$Laboratoire d'Astronomie Spatiale du CNRS, Traverse du Siphon, Les Trois Lucs, F-13012, Marseille, France \\ 
$^{**}$Dpt. of Physics \& Astronomy, JHU, 3400 N. Charles Street, Baltimore, MD 21218
}

\maketitle

\begin{abstract}
We present UV images of M~31 and M~32, as observed by HST
with the refurbished FOC.  The galaxies were observed through
the F175W and F275W filters, allowing the construction of
color magnitude diagrams (CMDs) for the hundreds of detected
sources found in each image.  Comparison of these data with the
stellar evolutionary tracks of horizontal branch stars and their
progeny shows that for the first time outside of our own Galaxy,
we may be measuring the colors of individual stars that are
evolving along post asymptotic giant branch (PAGB), post-early
AGB, and AGB-Manqu$\acute{\rm e}$ paths.  Searching to the 6$\sigma$ detection
limit, we find 1349 stars in M~31 and 183 stars in M~32.   We
compare the distribution of stars in the CMDs with the
expectations from theory. 

\end{abstract}

\section*{Motivation}

The spectra of elliptical and spiral galaxy bulges exhibit a
strong upturn shortward of 2700 \AA, dubbed the ``UV upturn.'' 
Characterized by the $m_{1550}-V$ color, the UV upturn shows strong
variation (ranging from 2.05--4.50 mag) in nearby quiescent
galaxies.  The UV upturn is thought to vary with the
composition of an evolved stellar population, because the
lifetimes of stars leaving the horizontal branch vary by orders of
magnitude.  Stars that evolve along the asymptotic giant branch
(AGB) and post-AGB evolutionary paths are short-lived in
comparison to the AGB-Manqu$\acute{\rm e}$ stars that spend 10$^7$ yr at high
luminosity (100 L$_{\odot}$) and high temperature (T$_{eff} > 30,000$K). 
The longer-lived stars are more efficient UV emitters, and thus
it is thought that they comprise a greater proportion of the
populations with the stronger UV upturns (see Brown et al.\ 1997 
for a review\cite{brown97}).
 
Many groups have tried to characterize the stellar populations in
early-type galaxies by fitting their composite spectral energy
distributions (SEDs) with the synthetic spectra of model stellar
populations (cf.\ Brown et al.\ 1997\cite{brown97}).  However, the considerable
ambiguity in this process has not determined unequivocally the
populations producing the UV flux.  Our goal in this work was
to directly characterize these stellar populations through UV
photometry of the individual stars in two galaxies: M~31 and
M~32.  To pursue this goal, we have taken deep exposures of
these galaxies with the F175W and F275W filters on the Faint
Object Camera (FOC).  The resulting color-magnitude diagrams
(CMDs) that can be constructed from this photometry can be
compared to the predictions of stellar evolutionary theory.
 
\section*{Observations}

The observations for this work are summarized in Table 1.  Our
photometry was performed with the IRAF package apphotx,
which we used to find stars that were detected at 6$\sigma$ above the
diffuse background.  This search found 1349 sources in M~31 and
183 in M~32.  We then simulated the data in order to 
determine our completeness limits and spurious source
contamination.  These simulations show that our detection 
capability drops below 50\% for stars dimmer than 24$^{th}$ 
magnitude in each individual filter for a given galaxy.  
However, our actual detection capability is somewhat better than that
estimated from these simulations,
since we generate our source list by searching in both filters.
Our magnitudes use
the STMAG system; $m_o$ in Table 1 corresponds to 1 count/sec in the filter.
We have constructed 3-color images of the galaxies (see Ferguson 
\& Brown, this volume) using our own FOC data and archival WFPC2
data. 

\begin{table}[h]
\caption{FOC Observations and Photometry}
\begin{tabular}{|l||c|c||c|c|}
\tableline
&\multicolumn{2}{c||}{M~31} & \multicolumn{2}{c|}{M~32}\\
\tableline
R.A.$_{2000}$&\multicolumn{2}{c||}{$0^h42^m44.74^s$}&\multicolumn{2}{c|}{$0^h42^m41.53^s$} \\
Dec.$_{2000}$ & \multicolumn{2}{c||} {$41^\circ 16\arcmin 8.04 \arcsec$} &\multicolumn{2}{c|}{$40^\circ 51\arcmin 51.82\arcsec$} \\
$D$ (kpc) & \multicolumn{2}{c||} {770} &  \multicolumn{2}{c|}{770} \\
$E(B-V)$ (mag) & \multicolumn{2}{c||} {0.11} & \multicolumn{2}{c|}{0.11} \\
$m_{1550}-V$ (mag)& \multicolumn{2}{c||}{3.51} & \multicolumn{2}{c|}{4.50} \\
\tableline
\tableline
& F175 & F275 & F175 & F275 \\
\tableline
Exp. (sec) & 19773 & 8390 & 16179 & 8990 \\
PSF FWHM (pix) & 4.6 & 4.0 & 4.3 & 3.9 \\
$m_o$ & 19.43 & 21.47 & 19.43 & 21.47 \\
\tableline
\end{tabular}
\end{table}

\begin{figure}[htp!]
\centerline{\epsfxsize=4.75in \epsffile{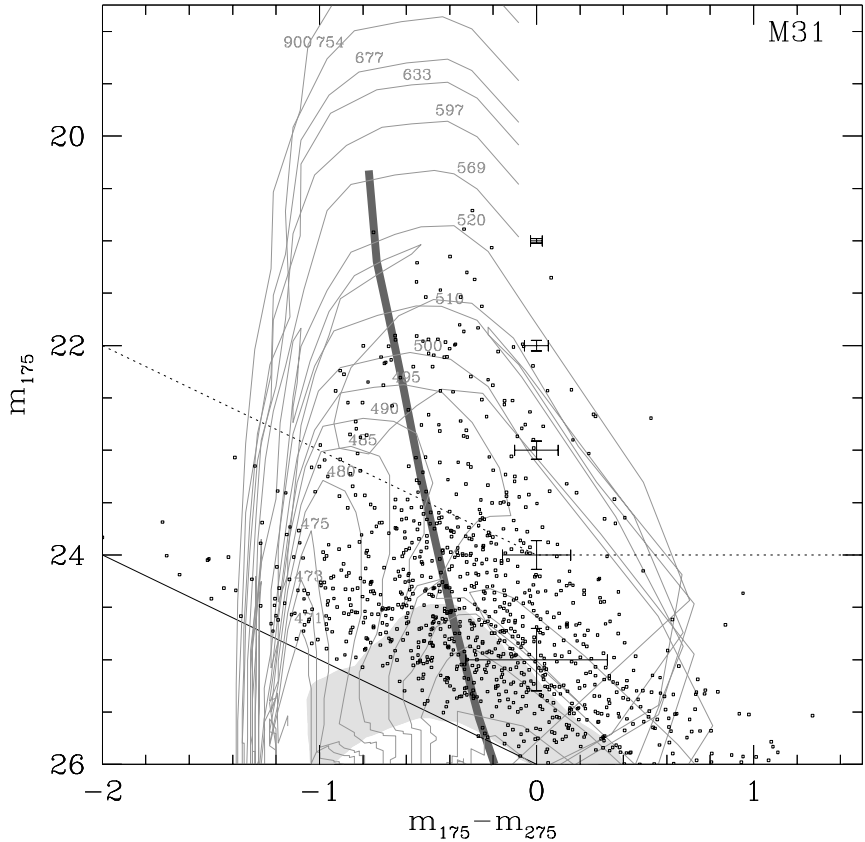}}
\vspace{15pt}
\caption{
A color-magnitude diagram (CMD) for the M~31 detected sources (black squares) 
brighter than 26$^{\rm th}$ magnitude (thin black line) in both colors.  
AGB-Manqu$\acute{\rm e}$ and post-early AGB stellar evolutionary tracks 
(thin grey curves; 0.471--0.520 M$_{\odot}$; 
Dorman et al.\ 1993\protect\cite{dorman93}) are labeled 
with 1000$\times \rm M_{ZAHB}$ in solar units.  Post-AGB evolutionary 
tracks (thin grey curves; 0.569--0.900 M$_{\odot}$; 
Vassiliadis \& Wood 1994\protect\cite{vassiliadis94}) are 
labeled with 1000$\times \rm M_{core}$ in solar
units; the AGB path is not shown for these tracks.  Stars should avoid the 
region where the evolution is relatively rapid (light grey) and cluster above 
the  region; instead, stars fill this 
``zone of avoidance.''    
If present, intermediate-mass (2.5--4 M$_{\odot}$) red giant branch stars
would cross the light grey region, but the much slower main sequence phase
(thick grey curve; 2.2--8 M$_{\odot}$) would be distinguishable if 
young stars
were populating this CMD.  The stars to the left (blueward) of all tracks
are most likely PAGB stars with planetary nebula; the PN emission lines which
would appear in the F175 filter (e.g., CIV $\lambda\lambda1548,1551$, 
CIII $\lambda1909$, HeII $\lambda1640$) would be strong enough
to cause the observed shifts relative to the tracks.
The thin dashed black line delineates our 50\% completeness limit for 
object detection in a single filter.  The error bars reflect our statistical
uncertainties at m$_{175}$=m$_{275}$=21, 22, 23, 24, and 25 mag; the
large uncertainties toward the bottom of the CMD may make it difficult
to detect the theoretical ``zone of avoidance.''}
\end{figure}
 
\begin{figure}[t!]
\centerline{\epsfxsize=4.75in \epsffile{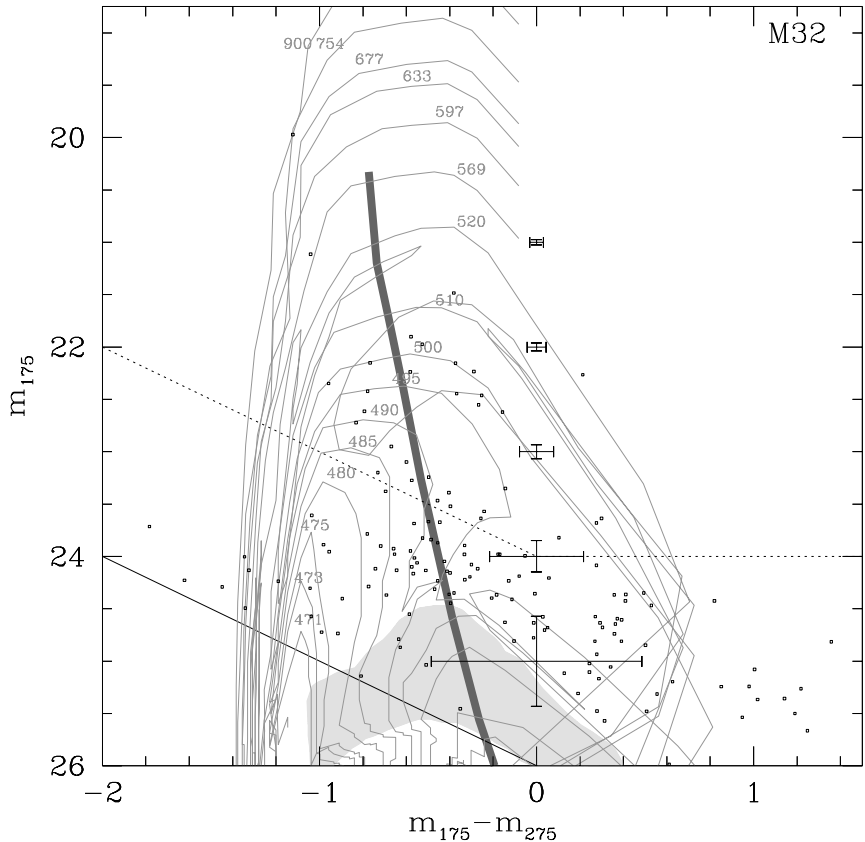}}
\vspace{15pt}
\caption{
A color-magnitude diagram (CMD) for the M~32 detected sources (black squares) 
brighter than 26$^{\rm th}$ magnitude (thin black line) in both colors (as
in Fig.~1 for M~31).  Since there are far fewer stars in this CMD (as compared
to Fig.~1), it is difficult to discern any clustering (or lack thereof).
However, it appears that the stars are again populating the light grey region.
}
\end{figure}

\section*{Comparison with Theory}

In Figs. 1 and 2, we have produced CMDs for these
sources, and compared them with stellar evolutionary tracks for
HB and post-HB stars.  The evolutionary tracks were placed in the CMDs by
finding the closest Kurucz (1993\cite{kurucz93}) synthetic spectrum for each 
step along
the track, normalizing it to agree with the distance and reddening for
the galaxies (Table 1), and then folding it through
the appropriate instrument response curves using the IRAF task calcphot.  
We have also performed Monte Carlo simulations of the
CMD that would arise from a population of extreme horizontal
branch (EHB) stars, chosen to reproduce the M~31 F175W countrate. 
In our simulations, this population is assumed to originate from a flat mass
distribution on the blue end of the HB, i.e., the number of stars
per unit mass is uniform across the EHB.  Our simulations show that
there should be a gap in the distribution of stars in each CMD (see
Figs. 1 \& 2).  The gap
is present because of the rapid evolution between two evolutionary
phases: the core He burning on the HB, and the later shell He burning in the
``slow blue phase.''

Although the exposure times on both galaxies are approximately equal, the
M~31 luminosity function peaks at $m_{275} = 26$ mag and $m_{175} = 25.5$ mag,
while the M~32 luminosity function peaks at $m_{275} = 25$ mag and 
$m_{175} = 24$ mag.  This is in line with theories of the UV upturn, since
one would expect brighter (and thus short-lived) stars in the galaxy with
the weaker UV upturn.

\section*{Puzzles}

Since the distribution of stars in our CMDs does not agree 
with the expectations from our simulations, we compared our data to that from
other investigations, in order to investigate possible problems with
our data.  These investigations turned up a few more discrepancies:

\begin{quote}
$\bullet$ The F175W countrate in both of our images is approximately
    60\% higher than that predicted from IUE observations 
    (private communication Calzetti) of M~31 and M~32. 
 
$\bullet$ Comparison of common sources in our data and that of pre-
    COSTAR F175W data (King et al.\ 1992\cite{king92}) shows our sources
    to be 1.1 mag fainter.

$\bullet$ Comparison of common sources in our data and that of archival 
WFPC2 F300W data shows that the mean $m_{275}-m_{300}$ color for these
sources is 0.23 mag, when it should be between 0.0 and -0.20 mag for 
hot UV sources at 10,000~K $\leq T_{eff} \leq$ 50,000~K.

\end{quote}
Because of these discrepancies, we cannot as yet characterize
the stellar populations in M~31 and M~32.  We are attempting
checks of the data calibration and our own calculations.  This
work is still in progress.

\end{document}